\documentclass[conference,10pt]{IEEEtran}
\IEEEoverridecommandlockouts
\usepackage{cite,mcite}
\usepackage{amsmath,amssymb,amsfonts,amsthm}
\usepackage{mathtools}
\usepackage{thmtools}
\usepackage{graphicx}
\usepackage{textcomp}
\usepackage{xcolor}
\usepackage{float}
\usepackage{pgfplots}
\usepackage{multirow}
\usepackage[hyphens]{url}

\newtheorem{definition}{Definition}[section]
\theoremstyle{definition}

\declaretheorem[style=remark,qed=$\blacktriangleleft$]{example}

\usetikzlibrary{arrows,positioning,shapes.geometric,decorations.pathmorphing}
\usetikzlibrary{pgfplots.fillbetween}
\usetikzlibrary{decorations.pathreplacing,decorations.markings}
\usetikzlibrary{plotmarks}
\pgfplotsset{compat=1.18}
\tikzset{->-/.style={decoration={
  markings,
  mark=at position #1 with {\arrow[scale=1.5,>=stealth]{>}}},postaction={decorate}}}

\begin{document}

\title{Generalized Spatially-Coupled Product-Like Codes Using Zipper Codes With Irregular Degree}

\author{\IEEEauthorblockN{Alvin Y. Sukmadji, Frank R. Kschischang, and Mohannad Shehadeh}
\IEEEauthorblockA{\textit{The Edward S. Rogers Sr. Department of Electrical \& Computer Engineering, University of Toronto}}
}

\maketitle

\begin{abstract}
Zipper codes with irregular variable degree are studied.
Two new interleaver maps---chevron and half-chevron---are described. Simulation results with shortened double-error-correcting
Bose--Chaudhuri--Hocquenghem constituent codes show that zipper codes with chevron and half-chevron interleaver maps outperform
staircase codes when the rate is below 0.86 and 0.91, respectively, at $10^{-8}$ output bit error rate operating point. In the
miscorrection-free decoding scheme, both zipper codes with chevron and half-chevron interleaver maps outperform staircase codes.
However, constituent decoder miscorrections induce additional performance gaps.
\end{abstract}

\section{Introduction}
In high-throughput communication systems such as optical transport networks and Ethernet, where
throughputs are on the order of hundreds of gigabits per second, it is imperative that the forward error correction (FEC) has low complexity while still
maintaining high data integrity. In optical transceivers, for example, it is known that about 22 to 35 percent of the power dissipation is due to the FEC
decoder \cite{morero,tucker1,tucker2}. To limit energy consumption, codes using hard-decision decoding, which consume about an order of magnitude less
power than those of soft-decision decoding, have been applied in this application \cite{justesen10,hu,yang,weiner,ou,lee13}.
Spatial coupling with iterative algebraic hard-decision decoding has been shown to perform well for various channel
conditions and parameters \cite{jian,lentmaier15}.

In this paper, we study zipper codes \cite{sukmadji} with irregular variable degree. This means that transmitted bits are copied
one or more times into the virtual buffer, causing each bit to be protected by two or more constituent codewords.
This generalizes many well-known spatially-coupled product-like codes, such as staircase
codes \cite{smith}, braided block codes \cite{feltstrom}, and oFEC \cite{humblet,g7093}, where each transmitted bit is protected by
two constituent codewords. We are motivated to consider irregular variable degree by analogy with irregular low-density parity check (LDPC)
codes, which are known to perform better than regular LDPC codes in the waterfall region \cite{luby,costello-scgldpc,mitchell}.

The rest of the paper is organized as follows. In Sec.~\ref{sec:zipper-codes}, we review the construction and
terminology used in zipper codes. The description of chevron and half-chevron interleaver maps, which achieve
regular and irregular degree distributions, respectively, is provided in Sec.~\ref{sec:degree-dist}.
Simulation results and discussions are given in Secs.~\ref{sec:design-example} and \ref{sec:discussion}, respectively.

Throughout this paper, let $\mathbb{N}=\{0,1,2,\ldots\}$ denote the set of all nonnegative integers,
and $[n]=\{0,1,\ldots,n-1\}$ for all $n\in\mathbb{N}\setminus\{0\}$.

\section{Zipper Codes}
\label{sec:zipper-codes}
In this section, we will review the construction and properties of zipper codes described in \cite{sukmadji}.

\subsection{Constituent Code, Zipping Pair, and Buffer}
Let $\mathcal{C}$ be a \emph{constituent code} over the binary field $\mathbb{F}_{2}=\{0,1\}$ of length $n$ and dimension $k$, and let
$r=n-k$ be the number of parity bits of $\mathcal{C}$. We assume that $\mathcal{C}$
is systematically encoded, i.e., the first $k$ bits are information bits
and the remaining $r$ bits are parity bits.

We fix a positive integer $m\leq k$ and partition $\mathbb{N}\times[n]$ into two disjoint subsets:
the \emph{virtual set} $A$ and the \emph{real set} $B$, with the property that
\[
  A = \mathbb{N}\times[m]\quad\text{and}\quad B=\mathbb{N}\times([n]\setminus[m]).
\]
We call $(A,B)$ a \emph{zipping pair}.

A \emph{buffer} $\mathbf{C}$ is a semi-infinite matrix with infinitely many rows and $n$ columns whose entries are in $\mathbb{F}_2$. For all
$i\in\mathbb{N}$ and $j\in[n]$, let $c_{i,j}$ denote the entry of $\mathbf{C}$ at the $i$th row and $j$th column.
Then, we call the sequence $\{(c_{i,0},\ldots,c_{i,m-1})\}_{i\in\mathbb{N}}$ to be the \emph{virtual buffer},
and similarly $\{(c_{i,m},\ldots,c_{i,n-1})\}_{i\in\mathbb{N}}$ to be the \emph{real buffer}. In other words,
the entries of the virtual buffer occupy the first $m$ columns of the buffer, and the entries of the real
buffer occupy the remaining $n-m$ columns. We therefore sometimes refer to $m$ as the \emph{virtual buffer width}
and $n-m$ as the \emph{real buffer width}.

\subsection{Interleaver Map}
An \emph{interleaver map} associated with a zipping pair $(A,B)$ is a function $\phi:A\rightarrow B$.
For each virtual position $(i,j)\in A$, the interleaver map gives a position in the real set $\phi(i,j)\in B$
from which to copy a symbol. In other words, for all $(i,j)\in A$, we have $c_{i,j}=c_{\phi(i,j)}$. For convenience,
we define $\phi_1(i,j)$ to be the row index of $\phi(i,j)$, i.e., if $\phi(i,j)=(i',j')$, then $\phi_1(i,j)=i'$.

We can also define the inverse image of the interleaver map. That is, for all $(i',j')\in B$, we have
\[
  \phi^{-1}(i',j')=\{(i,j)\in A:\phi(i,j)=(i',j')\}.
\]
The inverse image lists the positions in the virtual buffer in which the real bit at position $(i',j')$ is copied.
Furthermore, $\phi^{-1}_1(i',j')$ is defined similarly, i.e., if $\phi^{-1}(i',j')=\{(i_0,j_0),\ldots,(i_{\ell-1},j_{\ell-1})\}$,
then $\phi^{-1}_1(i',j')=\{i_0,\ldots,i_{\ell-1}\}$.

Two important, practical properties of an interleaver map are \emph{strict causality}
and \emph{periodicity}. The definitions are provided below.
\begin{definition}
An interleaver map $\phi$ is \emph{strictly causal} if for all $(i,j)\in A$, we have $\phi_1(i,j)< i$.
\end{definition}
\begin{definition}
An interleaver map $\phi$ is \emph{periodic} with period $\nu$ if for all $(i,j)\in A$, we have $\phi(i+\nu,j)=\phi(i,j)+(\nu,0)$.
\end{definition}
Throughout the rest of the paper, we assume that the interleaver maps are strictly
causal and periodic.

\subsection{Encoding and Decoding Procedure}
To encode row $i\geq 0$ of a zipper code we will follow the procedure below:
\begin{enumerate}
  \item Fill in $(c_{i,m},\ldots,c_{i,n-r-1})$ with new information bits.
  \item Fill in virtual bits $(c_{i,0},\ldots,c_{i,m-1})$ by copying bits from earlier rows according to
  the interleaver map, i.e., for all $j\in[m]$, $c_{i,j}=c_{\phi(i,j)}$.
  \item Compute the $r$ parity bits $(c_{i,n-r},\ldots,c_{i,n-1})$ using the systematic encoder of $\mathcal{C}$,
  thereby fulfilling the condition that
  $\mathbf{c}_i=(c_{i,0},\ldots,c_{i,n-1})\in\mathcal{C}$.
  \item Rows with index $i<0$ are initialized to be all-zero; these rows are not transmitted.
\end{enumerate}
Only the real bits $(c_{i,m},\ldots,c_{i,n-1})$ are transmitted over the channel. From this transmission scheme, we can
deduce that the \emph{rate} of the zipper code is $1-\frac{r}{n-m}$.

\begin{figure}[t]
\centering
\includegraphics{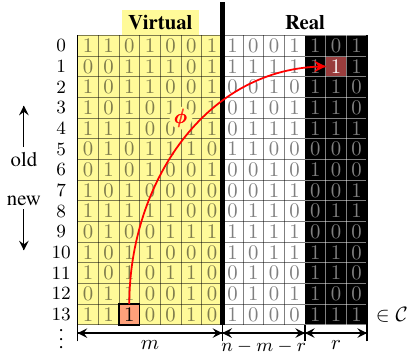}
\caption{Example of a zipper code with $n=14$, $m=7$, $r=3$. The first seven columns belong to the virtual buffer
and the last seven columns belong to the real buffer. The virtual and real buffers are demarcated by the thick
vertical line. The virtual bit in position $(13,2)$ is a copy of the bit in position $(1,12)$, i.e., $\phi(13,2)=(1,12)$.
This figure is modified from Fig.~1 of \protect\cite{sukmadji}.}
\label{fig:zipper}
\end{figure}

An example of a zipper code is shown in Fig.~\ref{fig:zipper}. Note that in this paper, we use a simplified construction
relative to that described in \cite{sukmadji}, namely, we assume a single constituent code $\mathcal{C}$
and identical virtual buffer width for all rows.

At the receiver, we use a sliding window decoder. The decoding
is performed on $M$ consecutive received constituent codewords $\mathbf{c}'_{i-M+1},\mathbf{c}'_{i-M+2},\ldots,\mathbf{c}'_{i}$.
The virtual part of each codeword is composed of copies of real bits from previously received (and partially corrected)
codewords. For each codeword in the decoding window, we attempt to decode the codeword using the constituent decoder of
$\mathcal{C}$. If the decoder declares an error at a certain position in the codeword, the decoder will flip the bit
at that position as well as all of its copies in the buffer. More concretely, if the decoder declares an error in position
$(i,j)\in A$, i.e., in the virtual buffer, then we correct by flipping the bits in positions
\[
\{\phi(i,j)\}\cup\phi^{-1}(\phi(i,j)).
\]
On the other hand, if the decoder declares an error in position $(i,j)\in B$, i.e., in the real buffer, we flip the bits in positions
\[
  \{(i,j)\}\cup\phi^{-1}(i,j).
\]
Once all rows in the decoding window are decoded, the decoder will repeat the process until no more corrections are
to be made, or when the maximum allowable number of iterations is reached. A few variants of the decoding procedure, including its
optimization, are described in \cite{sukmadji}.

\section{Specific Interleaver Maps}
\label{sec:degree-dist}

\subsection{Factor Graph}
The factor graph representation of a zipper code with zipping pair $(A,B)$ and interleaver map $\phi$ is a bipartite
graph $G=(V_b, V_c, E)$, where the two vertex sets are $V_b=B$, $V_c=\mathbb{N}$, and the edge set
\[
  E=\left\{((i,j),\ell):(i,j)\in B,~\ell\in\phi^{-1}_1(i,j)\cup\{i\}\right\}.
\]
The vertex set $V_{b}$ represents the bit positions in the real buffer, while $V_{c}$ represents the codeword (or row) indices.
The vertex $(i,j)$ joins vertex $\ell$ if and only if the real bit in position $(i,j)$ is involved in the
$\ell$th codeword, either as a real or virtual bit.

\begin{figure}[t]
\centering
\includegraphics{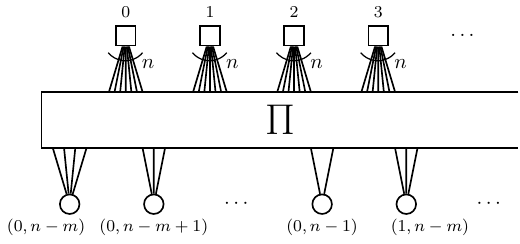}
\caption{Factor graph of zipper codes. Here, we have the degree of the real bit vertices (circle) $\deg_b(0,n-m)=4$,
$\deg_b(0,n-m+1)=3$, $\deg_b(0,n-1)=2$, and $\deg_b(1,n-m)=3$. The degree of each codeword/row
vertex (square) is $n$.}
\label{fig:factor-graph}
\end{figure}

The \emph{degree} of the vertex $(i,j)\in V_{b}$ is
\[
\deg_b(i,j)=1+|\phi^{-1}{(i,j)}|,
\]
where $|X|$ denotes the cardinality of the set $X$.
In other words, the degree of a real bit at a particular position represents the number of copies of that bit
in the buffer (including the real bit itself). The degree of vertex $\ell\in V_{c}$ is $n$ because each
constituent codeword is composed of $n$ bits. Fig.~\ref{fig:factor-graph} shows an example of a factor graph
of a zipper code.

\subsection{Regular and Irregular Interleaver Maps}
We provide a few examples of regular and irregular interleaver maps below.

\begin{example}[Staircase codes]
Staircase codes \cite{smith} are a type of zipper code with $n=2m$ for some positive integer $m$ and period $m$.
The interleaver map is defined by
\[
\phi(mi+r,j)=(m(i-1)+j,m+r)
\]
for all $i\in\mathbb{N},r\in[m],j\in[m]$.

\begin{figure}[t]
\centering
\includegraphics{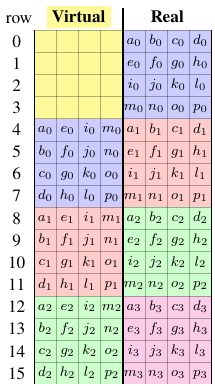}
\caption{Example of a staircase code (in its zipper code format) with $m=4$, $n=2m=8$.}
\label{fig:stc}
\end{figure}

Fig.~\ref{fig:stc} shows an example of a staircase code in its zipper code format.
From the figure, we can see that each real bit is only copied once in the virtual buffer, and so for all $(i,j)\in B$,
$\deg_b(i,j)=2$; thus, staircase codes are zipper codes that are regular with degree 2.
\end{example}
The examples presented in \cite{sukmadji} (braided block codes, tiled/delayed diagonal zipper codes, etc.)
are also examples of zipper codes that are regular with degree 2.

\begin{figure}[t]
\centering
\includegraphics{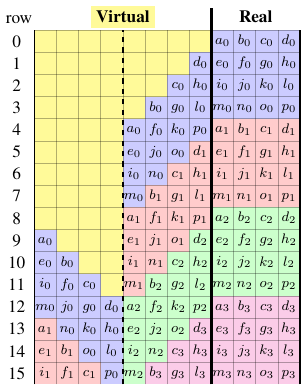}
\caption{Chevron interleaver map with $m'=4$, $m=2m'=8$, and $n=3m'=12$.}
\label{fig:chevron}
\end{figure}

\begin{example}[Chevron interleaver map]
In a chevron interleaver map, we have $m=2m'$ for some positive integer $m'$, and $n=3m'$.
The interleaver map for all $(i,j)\in A$ is as follows:
\begin{align*}
\phi(i,j) = \begin{cases}
(i-j-2m'-1,2m'+j)&\text{if}~j<m',\\
(i-2m'+j,m'+j)&\text{if}~m'\leq j<2m'.
\end{cases}
\end{align*}
An example of a zipper code with chevron interleaver map is shown in Fig.~\ref{fig:chevron}. Since each real bit is copied twice
in the virtual buffer, we have $\deg_b(i,j)=3$ for all $(i,j)\in B$.
Thus, zipper codes with the chevron interleaver map are regular with degree 3.
\end{example}

\begin{example}[Half-chevron interleaver map]
In a half-chevron interleaver map, we have $m=3m'$ and $n=5m'$ for some positive integer $m'$, with the following interleaver map:
\begin{align*}
\phi(i,j) = \begin{cases}
(i-j-4m'-1,3m'+j)&\text{if}~j<m',\\
(i-3m'+j,2m'+j)&\text{if}~m'\leq j<3m'.
\end{cases}
\end{align*}
An example of a zipper code with half-chevron interleaver map is shown in Fig.~\ref{fig:halfchevron}. One half of the real bits are
copied only once in the virtual buffer, while the other half are copied twice. Thus, zipper codes with half-chevron interleaver
map are irregular.

\begin{figure}[t]
\centering
\includegraphics{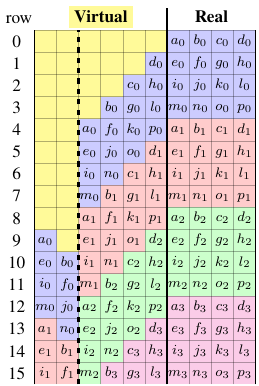}
\caption{Half-chevron interleaver map with $m'=2$, $m=3m'=6$, and $n=5m'=10$.}
\label{fig:halfchevron}
\end{figure}
\end{example}

\section{Design Example}
\label{sec:design-example}
\subsection{Simulation Results}
We performed Monte Carlo simulations over a binary symmetric channel of staircase codes and zipper codes with the chevron and
half-chevron interleaver maps with code rates
$0.75, 0.76,\ldots, 0.94$. All constituent codewords are shortened from the double-error-correcting ($t=2$)
Bose--Chaudhuri--Hocquenghem (BCH) $(1023,1003)$ code. We set the decoding window size to $8(n-m)$ consecutive rows with a maximum of
10 decoding rounds per window, and we shift the decoding window by $n-m$ rows for the next decoding sweep.
We use a few decoding techniques described in \cite{sukmadji}, namely exhaustive
decoding, fresh/stale flags, and periodic truncation.

\begin{figure}[t]
  \centering
  \includegraphics{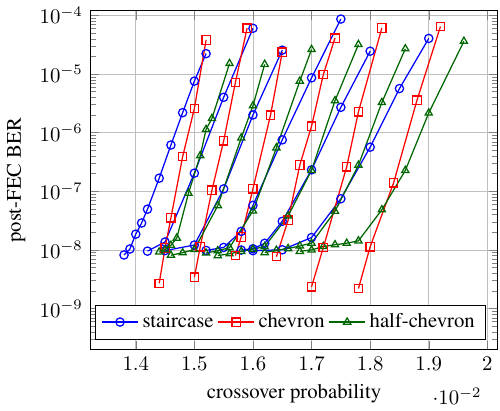}
  \caption{Monte Carlo simulation results of staircase codes and zipper codes with chevron and half-chevron interleaver maps of rate $0.75, 0.76,\ldots, 0.80$
  with $t=2$ constituent codes. For each type of interleaver map, the order in which each curve appears from left to right is for rate $0.80, 0.78,\ldots, 0.75$.}
  \label{fig:sim-result0}
\end{figure}

\begin{figure}[t]
  \centering
  \includegraphics{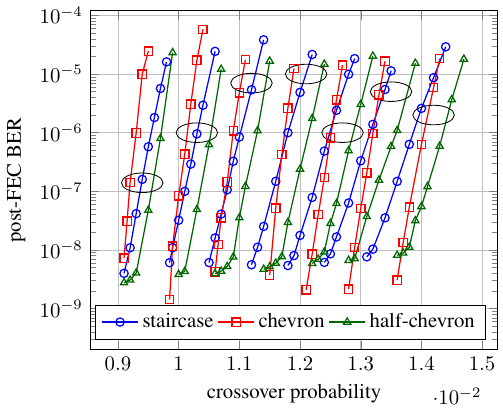}
  \caption{Monte Carlo simulation results of staircase codes and zipper codes with chevron and half-chevron interleaver maps of rate $0.81, 0.82,\ldots, 0.87$
  with $t=2$ constituent codes. For each type of interleaver map, the order in which each curve appears from left to right is for rate $0.87, 0.86,\ldots, 0.81$.}
  \label{fig:sim-result1}
\end{figure}

\begin{figure}[t]
  \centering
  \includegraphics{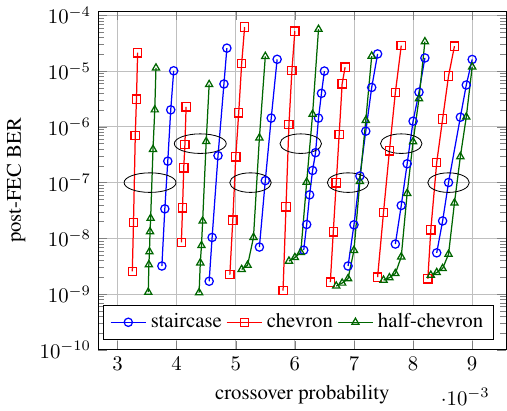}
  \caption{Monte Carlo simulation results of staircase codes and zipper codes with chevron and half-chevron interleaver maps of rate $0.88, 0.89,\ldots, 0.94$
  with $t=2$ constituent codes. For each type of interleaver map, the order in which each curve appears from left to right is for rate $0.94, 0.93,\ldots, 0.88$.}
  \label{fig:sim-result2}
\end{figure}

The Monte Carlo simulation results are summarized in Figs.~\ref{fig:sim-result0}--\ref{fig:sim-result2} and Table~\ref{tab:sim-result}. In
Table~\ref{tab:sim-result}, $\overline{m}=n-m$ denotes the real buffer width.
As a general characteristic, the chevron interleaver map has the highest slope, staircase has the smallest slope, and half-chevron is intermediate.
This implies that the curves will intersect, but the points of intersection depend on the code rate. Thus, no single interleaver map
is best in all cases. Further, some of the bit error rate (BER) plots for both staircase codes and zipper codes with
half-chevron interleaver map exhibit error floors at around $10^{-8}\sim10^{-9}$ post-FEC BER. However, no error floors are observed for zipper codes with
chevron interleaver map in that regime. Increasing the error-correcting radius to $t=3$ reduces the error floor for all the interleaver maps to a value below $10^{-9}$ (not measurable by our software implementation).

\begin{table}[ht]
\setlength{\tabcolsep}{3pt}
\centering
\caption{Simulation Parameters, $p^*$, and Gaps (in Decibels) to the Shannon Limit With $t=2$ Constituent Codes}
\begin{tabular}{|c|c|c|c|c|c|c|c|}
\hline
\multirow{2}{*}{Rate} & \multirow{2}{*}{$\overline{m}$} & \multicolumn{2}{|c|}{Stc ($n=2\overline{m}$)} & \multicolumn{2}{|c|}{Chev ($n=3\overline{m}$)} & \multicolumn{2}{|c|}{$\frac{1}{2}$Chev (\raisebox{-1.25ex}{\rule{0pt}{3.5ex}}$n=\frac{5}{2}\overline{m}$)}\\\cline{3-8}
 & & $p^{*}$ & Gap & $p^{*}$ & Gap & $p^{*}$ & Gap\\\hline
0.75 & 80 & 1.64e-02 & 1.819 & 1.80e-02 & 1.665 & 1.70e-02 & 1.760\\
0.76 & 84 & 1.58e-02 & 1.753 & 1.72e-02 & 1.616 & 1.64e-02 & 1.690\\
0.77 & 88 & 1.52e-02 & 1.684 & 1.64e-02 & 1.563 & 1.59e-02 & 1.614\\
0.78 & 92 & 1.46e-02 & 1.629 & 1.57e-02 & 1.505 & 1.54e-02 & 1.535\\
0.79 & 96 & 1.42e-02 & 1.533 & 1.51e-02 & 1.441 & 1.50e-02 & 1.454\\
0.80 & 100 & 1.39e-02 & 1.439 & 1.45e-02 & 1.372 & 1.45e-02 & 1.373\\\hline
0.81 & 106 & 1.32e-02 & 1.383 & 1.37e-02 & 1.327 & 1.38e-02 & 1.319\\
0.82 & 112 & 1.25e-02 & 1.323 & 1.29e-02 & 1.279 & 1.29e-02 & 1.274\\
0.83 & 118 & 1.19e-02 & 1.254 & 1.22e-02 & 1.219 & 1.24e-02 & 1.196\\
0.84 & 125 & 1.13e-02 & 1.191 & 1.15e-02 & 1.158 & 1.17e-02 & 1.142\\
0.85 & 134 & 1.06e-02 & 1.138 & 1.06e-02 & 1.126 & 1.09e-02 & 1.089\\
0.86 & 143 & 9.89e-03 & 1.075 & 9.89e-03 & 1.075 & 1.02e-02 & 1.037\\
0.87 & 154 & 9.19e-03 & 1.018 & 9.11e-03 & 1.030 & 9.37e-03 & 0.991\\\hline
0.88 & 167 & 8.45e-03 & 0.966 & 8.29e-03 & 0.991 & 8.59e-03 & 0.943\\
0.89 & 182 & 7.71e-03 & 0.913 & 7.46e-03 & 0.957 & 7.83e-03 & 0.894\\
0.90 & 200 & 6.97e-03 & 0.862 & 6.64e-03 & 0.922 & 7.02e-03 & 0.853\\
0.91 & 223 & 6.17e-03 & 0.821 & 5.83e-03 & 0.890 & 6.12e-03 & 0.832\\
0.92 & 250 & 5.41e-03 & 0.775 & 4.93e-03 & 0.884 & 5.30e-03 & 0.800\\
0.93 & 286 & 4.50e-03 & 0.768 & 4.08e-03 & 0.878 & 4.43e-03 & 0.786\\
0.94 & 334 & 3.77e-03 & 0.722 & 3.26e-03 & 0.878 & 3.56e-03 & 0.786\\\hline
\end{tabular}

\rule{0pt}{3ex}$p^*$ is the crossover probability such that the post-FEC BER is $10^{-8}$\\
 \label{tab:sim-result}
\end{table}

\subsection{Performance versus Rate Trade-off}
\begin{figure}[t]
\centering
\includegraphics{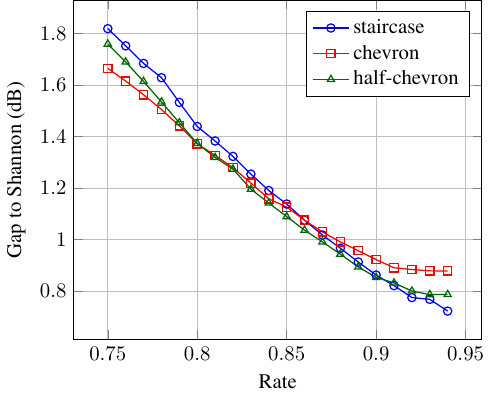}
\caption{Performance versus rate of staircase codes and zipper codes with chevron and half-chevron interleaver maps at $10^{-8}$ post-FEC BER.}
\label{fig:perf-rate}
\end{figure}

We plotted the performance versus rate of staircase codes and zipper codes with chevron and half-chevron interleaver maps in Fig.~\ref{fig:perf-rate}.
Here, \emph{performance} refers to the gap to the Shannon limit (see Sec.~IV.A. of \cite{sukmadji} for the definition) at $10^{-8}$ post-FEC BER operating point.
We observe that zipper codes with chevron interleaver map outperform the other two interleaver maps when the code rate $R$ is less than $0.80$.
On the other hand, when $0.80\leq R\leq 0.90$, zipper codes with half-chevron interleaver map outperform the others, and staircase codes
outperform the others when $R>0.90$.
However, Figs.~\ref{fig:sim-result0}--\ref{fig:sim-result2} suggest that these conclusions may change at a lower post-FEC BER, such as $10^{-15}$.

\section{Discussion}
\label{sec:discussion}
\subsection{Encoder Memory Size}
We define the \emph{encoder memory size} to be the maximum absolute difference between the current row and the row index from which to
copy the bit, i.e., the decoding memory size is
\[
 \max_{(i,j)\in A}|i-\phi_1(i,j)|.
\]
In our examples, given a real buffer width $(n-m)$, staircase codes have the smallest encoding memory size of $2(n-m)-1$,
followed by zipper codes with half-chevron interleaver map with $\frac{5}{2}(n-m)$, and chevron with $3(n-m)$. This implies
that in order to encode a particular row, both zipper codes with chevron and half-chevron interleaver maps must store more
bits compared to staircase codes in the encoder memory before those bits can be discarded.

\subsection{BCH Constituent Code Shortening and Miscorrection Rate}
One factor that causes zipper codes whose interleaver map induce real bits degree greater than two, such as chevron or half-chevron, to perform
worse than staircase codes, especially for high code rates, is the \emph{miscorrections} caused by the constituent decoder.
A miscorrection, i.e., a correction to a codeword that is not
transmitted, may happen when the Hamming weight of the error exceeds the decoding radius. In zipper codes, bit flips that are
caused by a constituent decoder miscorrection are copied over to other rows in the buffer, further deteriorating the decoding performance.

When the constituent codes are shortened, however, the bit positions of the BCH codeword that are not used (i.e., not transmitted) can be used as a
miscorrection detection mechanism. Specifically, suppose that a code of length $n$ is shortened from a BCH code of length $n_{\text{BCH}}$.
Without loss of generality, let $[n]$ be the bit indices used by the shortened code, and bits in positions $[n_{\text{BCH}}]\setminus[n]$ are not transmitted.
Suppose that the constituent decoder miscorrects and declares an error in a position contained in $[n_{\text{BCH}}]\setminus[n]$. The decoder will then
detect a miscorrection as an error is declared in the position of a bit that is not transmitted. Thus, in the case of an error weight that exceeds the
decoding radius, a miscorrection will happen if and only if all error locations that the decoder locates are contained in $[n]$.

\begin{figure}[t]
\centering
\includegraphics{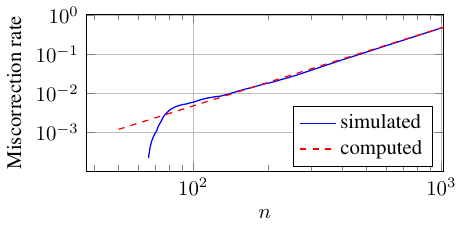}
\caption{Miscorrection rates of shortened $t=2$ BCH$(1023,1003)$ codes. Both simulated and computed \eqref{eq:miscrate} miscorrection rates are provided.}
\label{fig:misc-rate}
\end{figure}

A well-known estimate of the miscorrection rate of a $t$-error-correcting BCH (or Reed-Solomon) algebraic decoder given that more than $t$ symbol errors
have occurred is $1/t!$ \cite{mceliece}. We normalize the rate
by the probability that the $t$ error locations that the decoder declares are contained in $[n]$. Thus, an  estimate of the miscorrection rate of a
shortened BCH code of length $n$ shortened from length $n_{\text{BCH}}$, is given by
\begin{align}
  \frac{1}{t!}\cdot\frac{\binom{n}{t}}{\binom{n_{\text{BCH}}}{t}}\approx \frac{1}{t!}\left(\frac{n}{n_{\text{BCH}}}\right)^t.
  \label{eq:miscrate}
\end{align}
Fig.~\ref{fig:misc-rate} shows that the estimate of the miscorrection rate \eqref{eq:miscrate} of BCH codes shortened from $t=2$ BCH($1023,1003$)
agrees with the miscorrection
rate obtained from the Monte Carlo simulation for $n>150$. In general, the constituent codes for zipper codes with chevron
or half-chevron interleaver map are longer than those for staircase codes of the same rate. This implies that staircase codes have
stronger miscorrection detection capability than the other two.

\begin{figure}[t]
\centering
\includegraphics{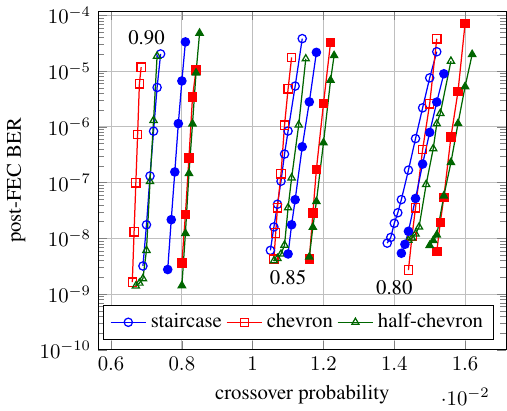}
\caption{Simulation results for zipper codes with rates 0.80, 0.85, and 0.90 using algebraic and miscorrection-free $t=2$ BCH constituent decoders.
Hollow and solid markers respectively denote the performance of algebraic and miscorrection-free decoders.}
\label{fig:sim-result-genie}
\end{figure}

Fig.~\ref{fig:sim-result-genie} and Table~\ref{tab:misc-gap} show the decoding performance of zipper codes with algebraic and miscorrection-free
$t=2$ BCH constituent decoders for rates 0.80, 0.85, and 0.90. In the miscorrection-free case, zipper codes with chevron and half-chevron interleaver maps
outperform staircase codes at $10^{-8}$ post-FEC BER. However, when algebraic decoders are used, the conclusions change due to the performance gaps induced
by the constituent decoder miscorrections. We can reduce the miscorrection rate by using anchor decoding \cite{hager} or soft-aided
decoding with marked bits \cite{lei}.

\begin{table}[t]
\setlength{\tabcolsep}{3pt}
\centering
\caption{Performance Gaps (in Decibels) Induced by Miscorrections at $10^{-8}$ Post-FEC BER}
\begin{tabular}{|c|c|c|c|c|c|c|}
\hline
\multirow{2}{*}{Rate} & \multicolumn{2}{|c|}{Staircase} & \multicolumn{2}{|c|}{Chevron} & \multicolumn{2}{|c|}{Half-Chevron}\\\cline{2-7}
 & Misc. Rate & Gap & Misc. Rate & Gap & Misc. Rate & Gap\\\hline
0.80 & 1.911e-02 & 0.051 & 4.300e-02 & 0.080 & 2.986e-02 & 0.070\\
0.85 & 3.432e-02 & 0.067 & 7.721e-02 & 0.131 & 5.362e-02 & 0.096\\
0.90 & 7.644e-02 & 0.122 & 1.720e-01 & 0.247 & 1.194e-01 & 0.184\\\hline
\end{tabular}
\label{tab:misc-gap}
\end{table}

\section{Conclusion}
In this paper, we have studied zipper codes with irregular variable degrees. Two interleaver maps are introduced: the chevron and half-chevron
interleaver maps. The former is a regular interleaver map with degree 3, while the latter is an irregular interleaver map with half
of the real bits having degree 2, while the other half having degree 3. Simulation results of the zipper codes with the two new interleaver maps and staircase codes
with $t=2$ shortened BCH constituent codes show that no single interleaver map is best. However, at $10^{-8}$ post-FEC BER, it is shown that
zipper codes with chevron and half-chevron interleaver maps outperform staircase codes at rates below 0.86 and 0.91, respectively.
In the miscorrection-free decoding scheme, we have shown that both zipper codes with chevron and half-chevron interleaver maps outperform staircase codes at
rates $0.75\sim0.90$. However, the miscorrection rates induce wider performance gaps for codes with longer constituent codes, such as zipper codes with chevron and half-chevron interleaver maps. This causes staircase codes to outperform the other two codes.

Future work may include designing a better interleaver map with irregular degrees, as well as analyzing the performance-complexity trade-offs and error
floor analysis of zipper codes with irregular degrees. We also consider using techniques described in \cite{hager} or \cite{lei} to reduce the constituent
decoder miscorrection rate.


\end{document}